\documentclass[10pt,conference]{IEEEtran}
\IEEEoverridecommandlockouts
\usepackage{cite}
\usepackage{amsmath,amssymb,amsfonts}
\usepackage{algorithmic}
\usepackage{fancyhdr}
\usepackage{graphicx}
\usepackage{textcomp}
\usepackage{xcolor}
\def\BibTeX{{\rm B\kern-.05em{\sc i\kern-.025em b}\kern-.08em
    T\kern-.1667em\lower.7ex\hbox{E}\kern-.125emX}}

\fancypagestyle{specialfooter}{%
\fancyhf{}

\fancyfoot[R]{ \noindent\fbox{%
\parbox{\textwidth}{%
{\footnotesize \bf © 2023 IEEE.  Personal use of this material is permitted.  Permission from IEEE must be obtained for all other uses, in any current or future media, including reprinting/republishing this material for advertising or promotional purposes, creating new collective works, for resale or redistribution to servers or lists, or reuse of any copyrighted component of this work in other works.}
}
}}
}

\begin{document}

\title{Quantum  negative sampling  strategy  for  knowledge graph embedding with   variational circuit}

\author{\IEEEauthorblockN{Pulak Ranjan Giri}
\IEEEauthorblockA{\textit{Quantum Computing Project} \\
\textit{KDDI Research, Inc.}\\
Fujimino-shi, Saitama, Japan \\
pu-giri@kddi-research.jp}
\and
\IEEEauthorblockN{Mori Kurokawa}
\IEEEauthorblockA{\textit{Quantum Computing Project} \\
\textit{KDDI Research, Inc.}\\
Fujimino-shi, Saitama, Japan \\
mo-kurokawa@kddi-research.jp}
\and
\IEEEauthorblockN{Kazuhiro Saito}
\IEEEauthorblockA{\textit{Quantum Computing Project} \\
\textit{KDDI Research, Inc.}\\
Fujimino-shi, Saitama, Japan \\
ku-saitou@kddi-research.jp}
}

\maketitle\thispagestyle{specialfooter}

\maketitle

\begin{abstract}
Knowledge graph  is   a collection of facts, known as triples(head, relation, tail), which  are  represented in   form of a network, where  nodes are  entities and edges are relations among the respective head and tail entities. Embedding of  knowledge graph for facilitating  downstream tasks such as  knowledge graph completion, link prediction, recommendation,   has been a  major area of research  recently in classical  machine learning.  Because the size of knowledge graphs are becoming larger, one of the   natural  choices  is  to  exploit quantum computing for knowledge graph embedding. Recently,    a hybrid quantum classical model  for knowledge graph embedding  has been  studied in which a variational  quantum circuit is trained.  
One of the important aspects in knowledge graph embedding is the sampling of negative triples, which plays a  crucial role in efficient training of the model.  In classical machine learning  various negative sampling strategies  have been studied.  In quantum knowledge graph embedding model, although we can use these strategies in principle, it is natural to ask  if we can exploit quantum advantage in negative sampling.    In this article we study such  a negative sampling strategy, which exploits quantum superposition,  and evaluate the model's performance with a knowledge graph database.

\end{abstract}


\section{Introduction} \label{in}
In recent years there has been huge interests  in the  study of knowledge graph(KG)  and its embedding \cite{wang}, due to  the  growing ability    to manipulate   large amount of data.    
Some of the extensively studied KGs \cite{lin}  include  UMLS, Kinship, Freebase,  YAGO and   Wordnet.

In order to use the knowledge graph for  downstream tasks such as  link predictions, subgraph classifications, question answering in AI  knowledge graph embedding(KGE)  have been very successful  in classical computing.

On the other hand, 
recently a quantum variational circuit model for knowledge graph embedding has been proposed  in  \cite{ma}, which is a generalisation of the  RESCAL model \cite{nickel} to the quantum domain.  Quantum model has  reduced complexity   \cite{ma} $\mathcal{O}(\mbox{poly}(\log d))$ compared   to the classical models' complexity   $\mathcal{O}(\mbox{poly}(d))$, where $d$ is the number of features. 

Besides the  choice of the variational circuit \cite{mori}, negative sampling strategies also play an important role in the  performance  of KGE model. 
In this article, we  propose a negative sampling strategy which exploits quantum superposition to form negative triples.

We   arrange  this article  in the following fashion:  A brief discussion  on  KGE  is given  in section    \ref{kg}.   In section \ref{vkge}   we discuss  knowledge graph embedding in hybrid  quantum classical  setting. In  sections    \ref{neg}  and \ref{expr}  negative sampling and its experimental results are  reported   and  finally in section \ref{con} we conclude with a discussion.

\section{Knowledge graph embedding} \label{kg}
In KGE  entities and relations  are encoded  as  vectors of  a low-dimensional  vector space.   Several KGE techniques \cite{wang} such as  RESCAL, TransE,  DistMult, TuckER, ComplEx,  have   been studied  in the literature.
Suppose  a typical element of a knowledge graph, called triple,  is    (head, relation, tail) =  $(h, r, t)$.   Then in the  RESCAL model $h$ and $t$ are encoded as vectors of a $d$-dimensional vector space and  relation as  a $d \times d$ matrix $M_r$.  The scoring function  $h^T M_r t$ then used to train the model.

\section{Variational KG embedding} \label{vkge}
In   variational quantum  circuit model for KGE   entities and relations of the KG are  embedded  in quantum circuits which are then optimised  by classical optimiser.  
Assume the entities  are   quantum states of the Hilbert space  $\mathcal{H}$  of dimensions $d= 2^n$. 
A variational quantum circuit for KGE  is given in Fig. 1. with $n=2$. 
A  fixed architecture  variational circuit  $\mathcal{U}(\alpha_{h_i})$ is  used to embed the head  as 
\begin{eqnarray}
 |h_i \rangle = \mathcal{U}(\alpha_{h_i})H^{\otimes n} |0 \rangle^{\otimes n}\,, 
\label{head}
\end{eqnarray}
where the set of $N_\alpha$  parameters $\alpha_{h_i} = \left[\alpha^1_{h_i}, \alpha^2_{h_i}, \cdots, \alpha^{N_\alpha}_{h_i}\right]$ for each head are optimised during training.  
The  embedded head  can be used as the tail  $ |t_i \rangle  =  |h_i \rangle $ as well.    Relations are embedded as the unitary matrices  $\mathcal{U}(\beta_{r_i})$, with 
the set of  $N_\beta$  parameters $\beta_{r_i} = \left[\beta^1_{r_i}, \beta^2_{r_i}, \cdots, \beta^{N_\beta}_{r_i}\right]$.     Head vectors  are evolved  by the relation operator  as 
\begin{eqnarray}
 |\tilde{h_{ij}} \rangle = \mathcal{U}(\beta_{r_i}) |h_j \rangle \,. 
\label{heade}
\end{eqnarray}
KGE model is trained by discriminating  the positive triples from the negative triples based on a well defined scoring function
\begin{figure}[h!]
  \centering
     \includegraphics[width=0.30\textwidth]{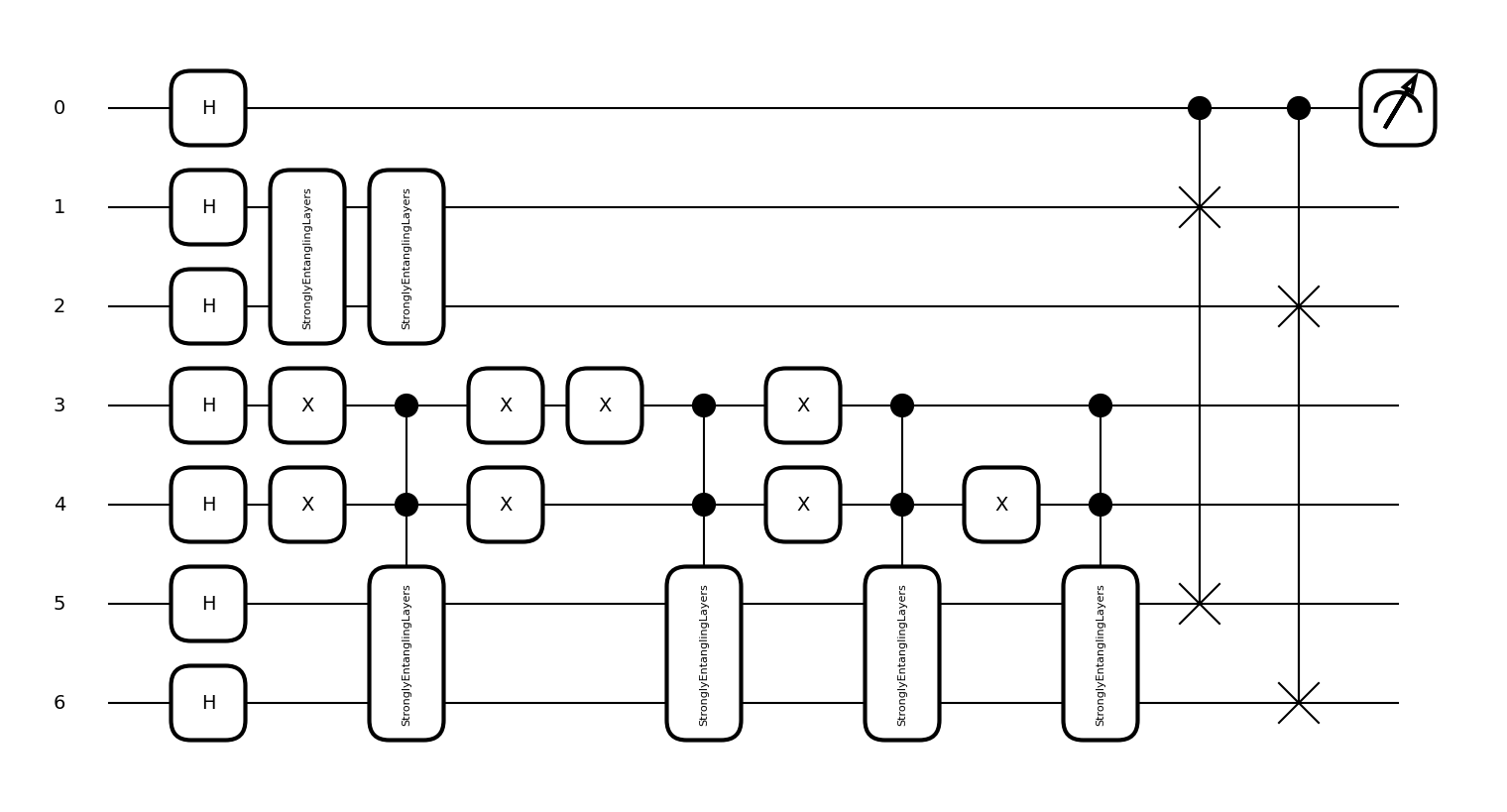}
          
       \caption{ A variational  quantum circuit for  knowledge graph embedding.  A fixed architecture quantum circuit  is  used  to embed  head and  tail,  and  another circuit is used to embed relation.}
\end{figure}
\begin{eqnarray}
\delta_{hrt} =  |\langle t | \mathcal{U}(\beta_{r}) |h \rangle|^2 \,.
\label{sc}
\end{eqnarray}
Over the course of training  period, evolution of the head  align itself  with the tail, i.e.,  $\delta_{h r t} \to 1$,  if the triple is  positive or align in orthogonal direction with respect to the tail, i.e.,  $\delta_{h r t} \to 0$,  if the triple is negative.  
Optimisation of  the variational circuit is performed by minimising  a suitably chosen  loss function, such as  mean square error   loss function: 
\begin{eqnarray}
L = \frac{1}{D} \sum_{h,r,t}\left(  \delta_{hrt} - y_{hrt} \right)^2 \,,
\label{loss}
\end{eqnarray}
where $D$ is the batch size for training and $y_{hrt}$ is the  label  corresponding to the triple $(h,r,t)$. 

\section{Negative sampling} \label{neg}
KGE   model works by discriminating positive triples from the  negative triples.  
It has been seen that the model performance  depends on how  and how many  negative  triples  are constructed   from a given positive triple \cite{qian}.  
Negative sampling strategies are broadly  divided into two categories: fixed sampling and dynamic sampling. 
Uniform    and Bernoulli sampling are  two widely used   fixed sampling strategy.  
And generative adversarial network(GAN) based     IGAN,  KBGAN   scheme are dynamic distribution negative sampling  strategies, which have been exploited recently for knowledge graph embedding for better performance.

We can use  the  above mentioned negative samplings in quantum variational circuit based  KGE.  However, as mentioned before, we here study a negative sampling, which is  based on the quantum superposition of multiple tails as shown in Fig. 1.  
The scoring function for the negative triple with four tails is  then given by 
\begin{eqnarray}
\delta_{hr \tilde{t}} =  \frac{1}{4} \sum_{i=1}^4 |\langle t_i | \mathcal{U}(\beta_{r}) |h \rangle|^2  \,,
\label{scs}
\end{eqnarray}
 which is used in the loss function for optimisation. 
\section{Experimental evaluations} \label{expr}
\subsection{Experimental setup:}  This experiment is  performed using  PennyLane's   ‘‘default.qubit" simulator.  Entities are embedded with  $4$ and $2$-qubits and 
additional $3$   ancilla qubits are required for our experiment.   Quantum circuit is optimised using  Adam optimiser  with   learning rate = $0.01$, loss function = mean square loss, epochs  $= 10$. PennyLane's  built-in quantum circuit,  StronglyEntanglingLayers, has been  used for entity embedding  and the same circuit with layer  $= 2$  has been used for relation embedding.  
\subsection{KG dataset:}  We have used Unified Medical Language Systems(UMLS) dataset, which has  entities = 135, relations = 46,  training triples = 5216, validation triples = 652,  test triples = 661. 
\subsection{Results:} 
For the evaluation of the performance of the  model we used link prediction. 
Percentage of  triples ranked up-to  1(Hits@1) and 10(Hits@10) are evaluated along with the  mean reciprocal  ratio(MRR)  and compared with the state-of-the-art results in Table 1. All the results are based on filtered data.   Although we used very small quantum circuit with  entity space dimension  = $4$ and $16$  for time limitation,  the accuracy of our  model is close to NeuralLP  results.  We can see  that as we increase entity embedding dimension from $2$-qubit to $4$-qubit the accuracy increase. Therefore,  further analysis  is necessary with higher  dimensional   embedding  of entity  and  with more expressive quantum circuits  with multiple layer  for  better comparison of  our model's performance with the state-of-the-art  performances. 

\begin{table}[htbp]
 \caption{Performance of our knowledge graph embedding model  compared with  state-of-the-art  models.} 

\begin{center}

    \begin{tabular}{ |  p{4.0cm }  | p{0.9cm } |  p{0.9cm} | p{0.9cm} |}
    \hline
    \textbf{Models}   & \textbf{MRR}  & \textbf{Hits@1} &  \textbf{Hits@10}  \\ \hline
    
    ConvE & $95.7$  & $93.2$  &  $99.4$  \\ \hline
    
    NeuralLP  &   $77.8$  & $64.3$  &  $96.2$   \\ \hline

    Our Model(4-qubit, 1 negative) & $74.3$  &  $63.0$   & $93.0$   \\  \hline

    Our Model(2-qubit, 4 negatives) & $59.0$  &  $43.4$   & $90.5$   \\  \hline
    
    Our Model(2-qubit, 3 negatives) & $60.9$  &  $46.9$   & $89.1$    \\  \hline

    Our Model(2-qubit, 2 negatives) & $59.0$  &  $44.5$   & $88.7$    \\  \hline
    
    Our Model(2-qubit, 1 negative) & $59.5$  &  $43.9$   & $87.6$    \\  \hline

 \end{tabular}
 \end{center}
 \end{table} 

\section{Conclusions} \label{con}
We studied a negative sampling strategy by making a quantum superposition of multiple tails and  evaluated its performance by training the  model with UMLS dataset.  


\section{Acknowledgements} 
This research was partially supported by JST CREST Grant Number JPMJCR21F2, Japan.




\end{document}